\documentclass{revtex4}
\usepackage{graphicx}
\usepackage{amsmath}
\DeclareMathOperator{\Tr}{Tr}
\DeclareMathOperator{\tr}{tr}
\DeclareMathOperator{\diag}{diag}
\DeclareMathOperator{\sgn}{sgn}
\DeclareMathOperator{\csch}{csch}
\begin{document}
\title{Dynamical Chiral Symmetry Breaking in NJL Model with a Constant External Magnetic Field}
\author{Song Shi$^{1,5}$}
\author{You-Chang Yang$^{2,3}$}
\author{Yong-Hui Xia$^{2}$}
\author{Zhu-Fang Cui$^{2,5}$}
\author{Xiao-Jun Liu$^{1}$}\email{liuxiaojun@nju.edu.cn}
\author{Hong-Shi Zong$^{2,4,5}$}\email{zonghs@nju.edu.cn}

\address{$^{1}$ Key Laboratory of Modern Acoustics, MOE, Institute of Acoustics, and Department of Physics, Collaborative Innovation Center of Advanced Microstructures, Nanjing University, Nanjing 210093, China}
\address{$^{2}$ Department of Physics, Nanjing University, Nanjing 210093, China}
\address{$^{3}$ School of Physics and Mechanical-Electrical Engineering, Zunyi Normal College, Zunyi 563002, China}
\address{$^{4}$ Joint Center for Particle, Nuclear Physics and Cosmology, Nanjing 210093, China}
\address{$^{5}$ State Key Laboratory of Theoretical Physics, Institute of Theoretical Physics, CAS, Beijing 100190, China}

\begin{abstract}

In this paper we develop a new method that is different from Schwinger proper time method to deduce the fermion propagator with a constant external magnetic field. In the NJL model, we use this method to find out the gap equation at zero and non-zero temperature, and give the numerical results and phase diagram between magnetic field and temperature. Beside these, we also introduce current mass to study the susceptibilities, because there is a new parameter (the strength of external magnetic field) in this problem, corresponding this new parameter, we have defined a new susceptibility $\chi_B$ to compare with the other two susceptibilities $\chi_c$ (chiral susceptibility) and $\chi_T$ (thermal susceptibility), and all of the three susceptibilities show than when current mass is not zero, the phase transition is a crossover, while for comparison, in the chiral limit, the susceptibilities show a second order phase transition. At last, we give out the critical coefficients of different susceptibilities in the chiral limit.

\bigskip


\end{abstract}
\maketitle
\section{Introduction}
In recent years, it is widely believed that strong magnetic field could play an important role in astrophysics \cite{duncan,duncan2} and high-energy physics \cite{sely,kha,kha2}. Especially in quantum field theory, many papers have shown that external magnetic field can affect dynamical chiral symmetry breaking (DCSB), which is also known as `Magnetic Catalysis' \cite{gusynin,gusynin2,ebert,shov,alan}. In the massless Nambu--Jona-Lasinio (NJL) model, with Schwinger proper time method \cite{alan,sch,klimenko,klimenko2} (one also can refer to Ref. \cite{ayala} for another method), it is shown that in both $1+3$ and $1+2$ dimensions a constant magnetic field can spontaneously break chiral symmetry no matter how small the coupling constant $G$ is, while in magnetic field free environment the chiral symmetry can be preserved when $G$ is smaller than a critical value $G_c$ \cite{klimenko,klimenko2,kle}.

From the magnetic catalysis effect, it is clear that the external magnetic field stimulates QCD condensation, with stronger magnetic field, the the value of QCD condensation or dynamic mass is bigger. On the other hand, as temperature participating in, the temperature weakens the condensation, with higher temperature, the value of QCD condensation or dynamic mass is smaller. The effect to condensation from magnetic field and temperature are opposite, hence putting these two conditions together can lead us to study their corporate contributions to the condensation.  Beside that, in the chiral limit, when the temperature is high enough (exceeding a critical value $T_c$), the system shall undergo a phase transition, the broke chiral symmetry will restore, and now with the presentation of external magnetic field, $T_c$ should be relative to the magnetic field, hence one of our purpose to study the influence of magnetic field and temperature to QCD condensation is to identify the function relation of magnetic field and critical temperature.

Susceptibilities and critical coefficients are important parameters to evaluate the chiral phase transition of QCD. In quantum field theory various susceptibilities are the linear responses of QCD condensation to various variables (e.g. temperature, current mass, chemical potential etc). In Ref. \cite{gupta}, it is found that the nonlinear susceptibilities are correlated with the cumulant of baryon-number fluctuations in experiments, and in Ref. \cite{zhao}'s work, the researchers have calculated these susceptibilities without magnetic field and established the correlation, but in their work this correlation still has some deviation from the experiments, since in the relativistic heavy-ion collisions there are the presence of strong magnetic field, we expect the magnetic field could amend such deviation, hence in this paper we study the susceptibilities in NJL model with magnetic field in order to support further work.

The following of this paper is organized in such a way: In the section II of this paper, we give the numerical results of dynamical mass dependence on magnetic field and temperature. In section III we study the susceptibilities with non-zero current mass, compare them to the chiral limit case, and discuss their phase transition properties. In this section we also calculate the critical coefficients for different susceptibilities. In Appendix A, we have proposed a new method that is different from Schwinger proper time, its result is equivalent to other methods', but when there is only external constant magnetic field rather than external electric field or external electromagnetic field, this new method is more convenient than the other method, at last of this appendix we discuss how this new method serves to thoroughly evaluate the contribution from infinitesimal imaginary term of the fermion propagator. In Appendix B, we give detailed deduction and discussion to the gap equation of two flavor NJL model with external magnetic field, because the electric charges of $u$ $d$ quarks are different, we need to prove the gap equation's self-consistency with rigorous in this appendix.

\section{The Gap Equations and Numerical Results in Chiral Limit}
In $1+3$ dimensions, the bosonized two flavor NJL lagrangian with external magnetic field is
\begin{equation}
\mathcal{L}=\bar\psi(i/\kern-0.5em\partial+e/\kern-0.65em A\otimes Q-\sigma-i\gamma^5\otimes\vec\pi\cdot\vec \tau)\psi-\frac{N_c}{2G}\Sigma^2,\label{L}
\end{equation}
\begin{equation}
Q=\left(\begin{array}{cc}q_u&0\\0&q_d\end{array}\right),\quad q_u=\frac{2}{3},\quad q_d=-\frac{1}{3},\quad q_f=q_u,q_d,
\end{equation}
\begin{equation}
\Sigma^2=\sigma^2+\pi^2,\quad\pi^2\equiv|\vec\pi|^2=\sum_{i=1}^3\pi_i^2,
\end{equation}
\begin{equation}
\vec \tau=(\tau^1,\tau^2,\tau^3),\quad \tau^1=\left(\begin{array}{cc}0&1\\1&0\end{array}\right),\quad \tau^2=\left(\begin{array}{cc}0&-i\\i&0\end{array}\right),\quad \tau^3=\left(\begin{array}{cc}1&0\\0&-1\end{array}\right),\quad\vec\pi\cdot\vec \tau=\pi_i\tau^i.
\end{equation}
Here we have already assumed the current mass of fermion is zero, and $A_\mu$ is the potential of external magnetic field, in order to serve the purpose of a constant external magnetic field, $A_\mu$ can be defined as
\begin{equation}
(A_0,A_1,A_2,A_3)=(0,\frac{B}{2}x^2,-\frac{B}{2}x^1,0),
\end{equation}
$x^1$ and $x^2$ are the second and the the third component of time-space coordinates $(x^0,x^1,x^2,x^3)$.
From Eq. (\ref{L}), the fermion propagator of different flavor with magnetic field should be
\begin{equation}
\hat S_f=\frac{1}{\gamma^\mu\hat\Pi^f_\mu-\sigma},\qquad\hat\Pi^f_\mu=i\partial_\mu+q_feA_\mu,
\end{equation}
throughout this paper, the pion condensation is thought to be zero ($\vec\pi=0$), because its existence would violate parity, a detailed analysis is discussed in Appendix B.

A more practical version of the fermion propagator is
\begin{equation}
\hat S_f=\frac{/\kern-0.6em\hat\Pi^f+\sigma}{(/\kern-0.6em\hat\Pi^f)^2-\sigma^2}=\frac{/\kern-0.6em\hat\Pi^f+\sigma}{(\hat\Pi^f)^2-\sigma^2-q_feB\sigma^{12}},\label{propagator}
\end{equation}
\begin{equation}
\sigma^{12}=\diag(1,-1,1,-1).
\end{equation}

In Eq. (\ref{propagator}), rather than employing Schwinger proper time method to rewrite the denominator of $\hat S_f$, we propose a new method to deal with this denominator. The detailed deduction of this new method has shown in Appendix A, here we only give the final result.

First of all, the general gap equation in two flavor NJL model with zero temperature is (the rigorous deduction of this equation is shown in Appendix B)
\begin{equation}
\frac{\sigma}{G}\int d^4x=i\sum_f\Tr\hat S_f.\label{gap}
\end{equation}
Now we can employ the results from Appendix A, replace Eq. (\ref{trace2}) into Eq. (\ref{gap}), and simplify the gap equation, we have
\begin{equation}
\frac{4\pi^2}{G}=\sum_f|q_f|eB\int_0^{+\infty}\frac{e^{-\sigma^2s}}{s}\coth(|q_f|eBs)\,ds,\label{gap0}
\end{equation}
and by making a cutoff $\frac{1}{\Lambda^2}$ to the low limit of $s$'s integral, eventually we get the gap equation that is suitable for numerical calculation,
\begin{equation}
\frac{4\pi^2}{G}=N_f\int_{\frac{1}{\Lambda^2}}^{+\infty}\frac{e^{-\sigma^2s}}{s^2}\,ds+\sum_f|q_f|eB\int_0^{+\infty}\frac{e^{-\sigma^2s}}{s}\bigg[\coth(|q_f|eBs)-\frac{1}{|q_f|eBs}\bigg]\,ds.\label{gap2}
\end{equation}
In Eq. (\ref{gap2}), $N_f=2$, and the divergent part and convergent part are separated, hence the cutoff only affects the divergent part.

At the finite temperature, the gap equation is slightly different with Eq. (\ref{gap}), which is
\begin{equation}
\frac{\sigma}{G}\int_0^\beta d\tau\int d^4x=-\sum_f\Tr\hat S_f,\qquad\beta=\frac{1}{T},\label{gapt}
\end{equation}
and comparing with Eq. (\ref{propagator}), the $\hat p_0$ operator has changed into $\frac{\partial}{\partial\tau}$, with an appropriate eigenstate, there is
\begin{equation}
\frac{\partial}{\partial\tau}|m\rangle_0=i\omega_m|m\rangle_0,\quad\omega_m=(2m+1)\pi T,\quad m\in\{\cdots,-2,-1,0,1,2,\cdots\},
\end{equation}
then we can deduce an intermediate gap equation,
\begin{equation}
\frac{4\pi^2}{G}=2T\sum_f|q_f|eB\int d\Pi_3\sum_{m=-\infty}^{+\infty}\sum_{n=0}^{+\infty}\frac{(2-\delta_{0n})}{\omega_m^2+2n|q_f|eB+\Pi_3^2+\sigma^2}.
\end{equation}
In order to inherit the same cutoff $\frac{1}{\Lambda}$ from Eq. (\ref{trace2}), we choose to sum all of $2n|q_f|eB$ up rather than add $\omega_m$ up,
\begin{eqnarray}
\frac{4\pi^2}{G}&=&2T\sum_f|q_f|eB\int d\Pi_3\sum_{m=-\infty}^{+\infty}\sum_{n=0}^{+\infty}(2-\delta_{0n})\int_0^{+\infty}e^{-(\omega_m^2+2n|q_f|eB+\Pi_3^2+\sigma^2)s}\,ds\nonumber\\
&=&4\sqrt{\pi}T\sum_f|q_f|eB\int_0^{+\infty}\bigg(\sum_{m=0}^{+\infty}e^{-\omega_m^2s}\bigg)\frac{e^{-\sigma^2s}}{\sqrt{s}}\coth(|q_f|eBs)\,ds,\label{gapt0}
\end{eqnarray}
introducing the cutoff,
\begin{eqnarray}
\frac{4\pi^2}{G}&=&4\sqrt{\pi}N_fT\int_\frac{1}{\Lambda^2}^{+\infty}\bigg(\sum_{m=0}^{+\infty}e^{-\omega_m^2s}\bigg)\frac{e^{-\sigma^2s}}{s\sqrt{s}}\,ds\nonumber\\&&
+4\sqrt{\pi}T\sum_f|q_f|eB\int_0^{+\infty}\bigg(\sum_{m=0}^{+\infty}e^{-\omega_m^2s}\bigg)\frac{e^{-\sigma^2s}}{\sqrt{s}}\bigg[\coth(|q_f|eBs)-\frac{1}{|q_f|eBs}\bigg]\,ds.\label{gap3}
\end{eqnarray}

Now we can use Eq. (\ref{gap2}) to identify the relation between dynamical mass of NJL and constant external magnetic field $eB$, referring to Ref. \cite{inagaki}, for $f_\pi=93\text{MeV}$, $m_\pi=138\text{MeV}$ and the current mass $m_0=5.5\text{MeV}$, the value of cutoff $\Lambda$ and $G$ are
\begin{equation}
\Lambda=0.991\text{GeV},\qquad G=25.4\text{GeV}^{-2}.
\end{equation}
\begin{figure}
\centering
\includegraphics[width=3in]{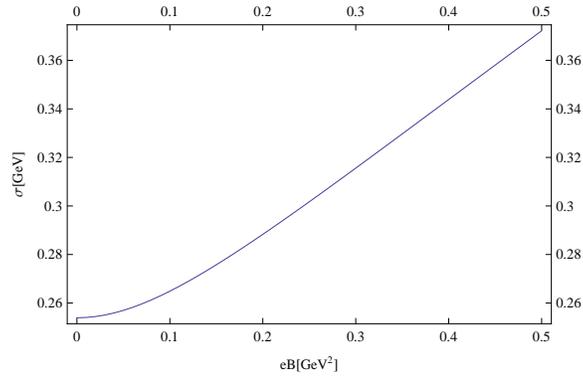}
\caption{The $eB$ dependance of dynamical mass $\sigma$ when zero temperature. When $eB=0$, $\sigma=0.25\text{GeV}$.\label{i}}
\end{figure}
Having set up these values, we are able to draw the $\sigma$-$eB$ relation in Fig. \ref{i}, it is clear that when the strength of magnetic field increases, the dynamical mass will increases with it, the chiral symmetry is always broken, and when the magnetic field is strong enough, the dynamical mass has a nearly linear response to the magnetic field.
\begin{figure}
\centering
\includegraphics[width=3in]{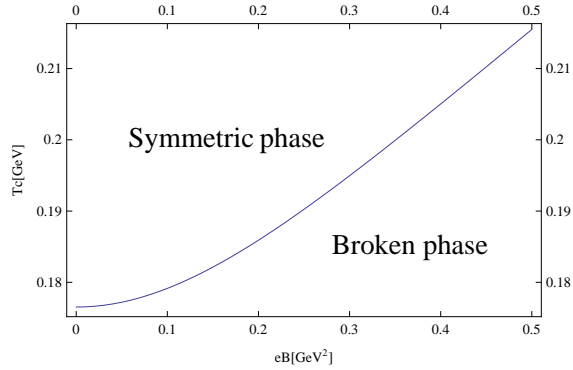}
\caption{The $eB$ dependance of critical temperature $T_c$. When $eB=0$, $T_c=0.17\text{GeV}$\label{ii}}
\end{figure}
\begin{figure}
\centering
\includegraphics[width=3in]{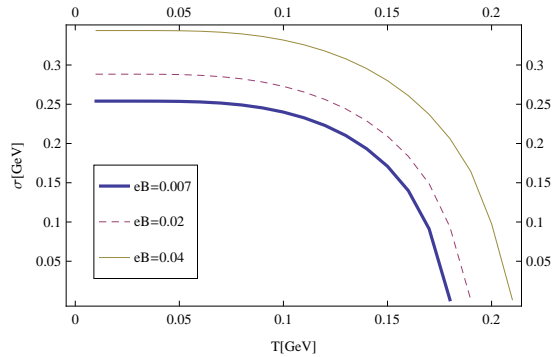}
\caption{The temperature dependance of dynamical mass $\sigma$ with different fixed $eB$.\label{iii}}
\end{figure}
\begin{figure}
\centering
\includegraphics[width=3in]{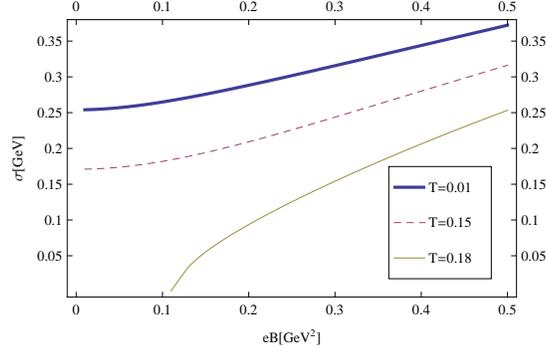}
\caption{The $eB$ dependance of dynamical mass $\sigma$ with different fixed temperatures.\label{iv}}
\end{figure}

Actually, when magnetic field is strong enough, the gap equation can be simplified to
\begin{equation}
\frac{4\pi^2}{G}=\sum_fq_feB\int_{\frac{1}{\Lambda^2}}^{+\infty}\frac{e^{-\sigma s}}{s}\,ds.\label{gapsimple}
\end{equation}
Because in Eq. (\ref{gap0}), when the cutoff $1/\Lambda^2$ of integral variable $s$ has made, there is
\begin{equation}
\lim_{eB\to+\infty}\coth(|q_f|eBs)=1,\quad s\ge\frac{1}{\Lambda^2}.
\end{equation}
Of cause, Eq. (\ref{gapsimple}) can also be deduced from Eq. (\ref{trace}) by throwing away the summation of $n$ only leaving the $n=0$ term along.

When temperature is not zero, it turns out there is a critical temperature $T_c$, as long as the system's temperature exceeds the critical temperature, the dynamical mass will be zero, which corresponds to the restoration of chiral symmetry. But for different magnetic field, the critical temperatures are different, as shown in Fig. \ref{ii}, $T_c$ is increasing along with $eB$'s increase. While in Fig. \ref{iii} and Fig. \ref{iv} they give us magnetic field or temperature dependance of dynamical mass. In Fig. \ref{iii}, the bigger the strength of $eB$, the upper the $\sigma\text{-}T$ curves, while in Fig. \ref{iv}, the higher the temperature, the lower the $\sigma\text{-}eB$ curves, and in this figure, the curve $T=0.2\text{GeV}$ is a little different with the other two, because in this situation, the magnetic field need to exceed a specific quantity to produce dynamical mass, while below this specific quantity, the chiral symmetry is unbroken.

In Fig. \ref{iii}, at different $eB$, the $\sigma$-$T$ curves are separating with each other, this a solid evidence that in NJL mean field approximation, there will have no inverse magnetic catalysis, and here we also can use a simple mathematics analysis to support this conclusion. Taking the cut off Eq. (\ref{gapt0}) for discussion, $eB\coth(|q_f|eBs)$ is a monotone increasing function of $eB$, hence the RHS of Eq. (\ref{gapt0}) is increasing along with $eB$ increasing, this will obviously cause the $\sigma$ monotonously increasing whichever the temperature is.

\section{Susceptibilities and Critical Coefficients}
In this section, we refer to Ref. \cite{du,yin}'s works, use the known gap equation to study various susceptibilities. Because we do not consider the chemical potential in this article, there will be no susceptibilities related to chemical potential, such as quark number susceptibility $\chi_q$ and vector-scalar susceptibility $\chi_{vs}$, but after introducing external magnetic field in NJL model, we can define a new susceptibility related to $eB$, named as magnetic field susceptibility $\chi_B$. Therefore in this section, we are going to study three kinds of susceptibilities, $\chi_B$, chiral susceptibility $\chi_c$ and thermal susceptibility $\chi_T$.

In order to study these susceptibilities, firstly, we need to enhance the NJL model Eq. (\ref{L}) with current mass $m$, correspondingly, the new gap equation can be simply achieved by replacing $\sigma$ with $\sigma+m$ at the RHS of Eq. (\ref{gap}) and Eq. (\ref{gapt}). We assume the cutoff and coupling constant is current mass irrelevant, hence the intermediate and final gap equation are
\begin{equation}
\frac{4\pi^2}{G}\sigma=4\sqrt{\pi}T(\sigma+m)\sum_f|q_f|eB\int_0^{+\infty}\bigg(\sum_{m=0}^{+\infty}e^{-\omega_m^2s}\bigg)\frac{e^{-(\sigma+m)^2s}}{\sqrt{s}}\coth(|q_f|eBs)\,ds,\label{gapt2}
\end{equation}
\begin{eqnarray}
\frac{4\pi^2}{G}\sigma&=&4\sqrt{\pi}N_fT(\sigma+m)\int_\frac{1}{\Lambda^2}^{+\infty}\bigg(\sum_{m=0}^{+\infty}e^{-\omega_m^2s}\bigg)\frac{e^{-(\sigma+m)^2s}}{s\sqrt{s}}\,ds\nonumber\\&&
+4\sqrt{\pi}T(\sigma+m)\sum_f|q_f|eB\int_0^{+\infty}\bigg(\sum_{m=0}^{+\infty}e^{-\omega_m^2s}\bigg)\frac{e^{-(\sigma+m)^2s}}{\sqrt{s}}\bigg[\coth(|q_f|eBs)-\frac{1}{|q_f|eBs}\bigg]\,ds.\label{gapt3}
\end{eqnarray}
Even with current mass, the properties of dynamic mass $\sigma$ are qualitatively similar to Fig. \ref{i}-\ref{iv}.

Because $\langle\bar\psi\psi\rangle\propto(-\sigma)$, and we have assumed the coupling constant $G$ and cutoff $\Lambda$ are independent of current mass, temperature and magnetic field et cetera, therefore in this article, we define the susceptibilities as
\begin{equation}
\chi_c=\frac{\partial\sigma}{\partial m},\quad\chi_T=-\frac{\partial\sigma}{\partial T},\quad\chi_B=\frac{\partial\sigma}{\partial(eB)}.
\end{equation}
In Eq. (\ref{gapt3}), treating $\sigma$ as the implicit function of $m$, $T$ and $eB$, $\sigma=\sigma(m,T,eB)$, we can make partial differentiations of $m$, $T$ and $eB$ and get the corresponding equations for the susceptibilities.

For ensuring the equations are easy to read and analyse, firstly we define a few functions to represent some complicate formulas,
\begin{eqnarray}
f_m(\sigma,m,T,eB)&=&8\sqrt{\pi}N_fT\int_\frac{1}{\Lambda^2}^{+\infty}\bigg(\sum_{m=0}^{+\infty}e^{-\omega_m^2s}\bigg)\frac{e^{-(\sigma+m)^2s}}{\sqrt{s}}\,ds\nonumber\\&&
+8\sqrt{\pi}T\sum_f|q_f|eB\int_0^{+\infty}\bigg(\sum_{m=0}^{+\infty}e^{-\omega_m^2s}\bigg)e^{-(\sigma+m)^2s}\sqrt{s}\bigg[\coth(|q_f|eBs)-\frac{1}{|q_f|eBs}\bigg]\,ds,\label{fm}
\end{eqnarray}
\begin{eqnarray}
f_T(\sigma,m,T,eB)&=&8\pi^\frac{5}{2}N_fT^2\int_\frac{1}{\Lambda^2}^{+\infty}\bigg(\sum_{m=0}^{+\infty}(2m+1)^2e^{-\omega_m^2s}\bigg)\frac{e^{-(\sigma+m)^2s}}{\sqrt{s}}\,ds\nonumber\\&&
+8\pi^\frac{5}{2}T^2\sum_f|q_f|eB\int_0^{+\infty}\bigg(\sum_{m=0}^{+\infty}(2m+1)^2e^{-\omega_m^2s}\bigg)e^{-(\sigma+m)^2s}\sqrt{s}\bigg[\coth(|q_f|eBs)-\frac{1}{|q_f|eBs}\bigg]\,ds,\nonumber
\\\label{ft}
\end{eqnarray}
\begin{equation}
f_B(\sigma,m,T,eB)=4\sqrt{\pi}T\sum_f|q_f|\int_0^{+\infty}\bigg(\sum_{m=0}^{+\infty}e^{-\omega_m^2s}\bigg)\frac{e^{-(\sigma+m)^2s}}{\sqrt{s}}
[\coth(|q_f|eBs)-(|q_f|eBs)\csch^2(|q_f|eBs)]\,ds,\label{fb}
\end{equation}
although these formulas are complicate, but they are all positive.

Now by employing Eq. (\ref{gapt3}), (\ref{fm}), (\ref{ft}) and (\ref{fb}), we are able to deduce the equations for susceptibilities,
\begin{equation}
\chi_c=\frac{1}{\frac{m}{\sigma+m}+\frac{G}{4\pi^2}(\sigma+m)^2f_m}-1,\label{chic}
\end{equation}
\begin{equation}
\chi_T=(1+\chi_c)\bigg[\frac{G}{4\pi^2}(\sigma+m)f_T-\frac{\sigma}{T}\bigg],\quad\text{or}\quad\chi_T=-\frac{G}{4\pi^2}(1+\chi_c)\bigg[\frac{\partial(\text{RHS})}{\partial T}\bigg]_\sigma,\label{chit}
\end{equation}
\begin{equation}
\chi_B=\frac{G}{4\pi^2}(1+\chi_c)(\sigma+m)f_B.\label{chib}
\end{equation}
For convenience of numerical calculation, Eq. (\ref{chit}) provides a second formula to calculate thermal susceptibility, in this formula, $(\text{RHS})$ represents the RHS of Eq. (\ref{gapt3}), and the partial differentiation of $(\text{RHS})$ only operates on parameter $T$ but treats $\sigma$ as a constant.

\begin{figure}
\centering
\includegraphics[width=3in]{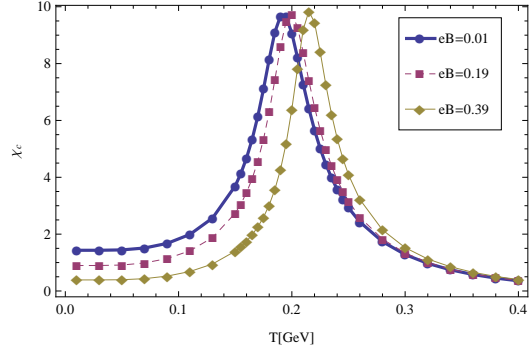}
\caption{The temperature dependance of $\chi_c$ with different $eB$.\label{ci}}
\end{figure}
\begin{figure}
\centering
\includegraphics[width=3in]{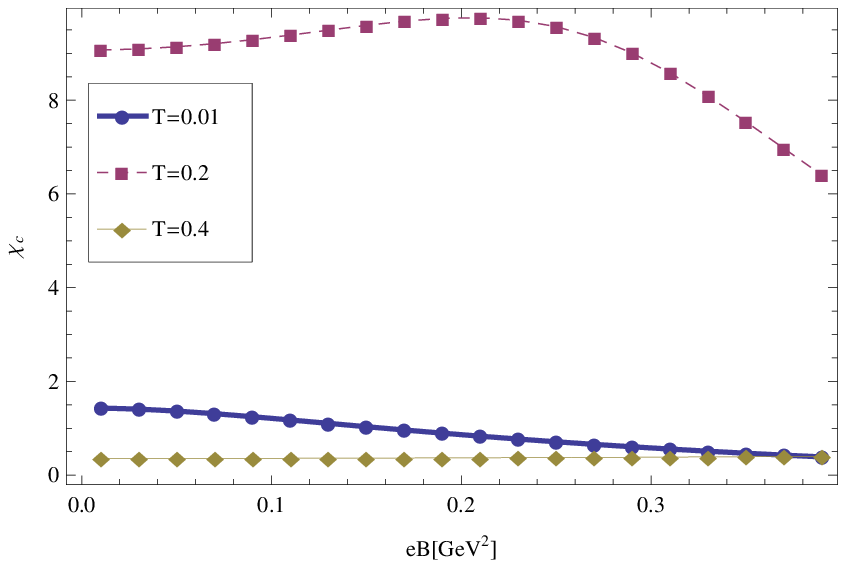}
\caption{The $eB$ dependance of $\chi_c$ with different temperature.\label{cii}}
\end{figure}
\begin{figure}
\centering
\includegraphics[width=3in]{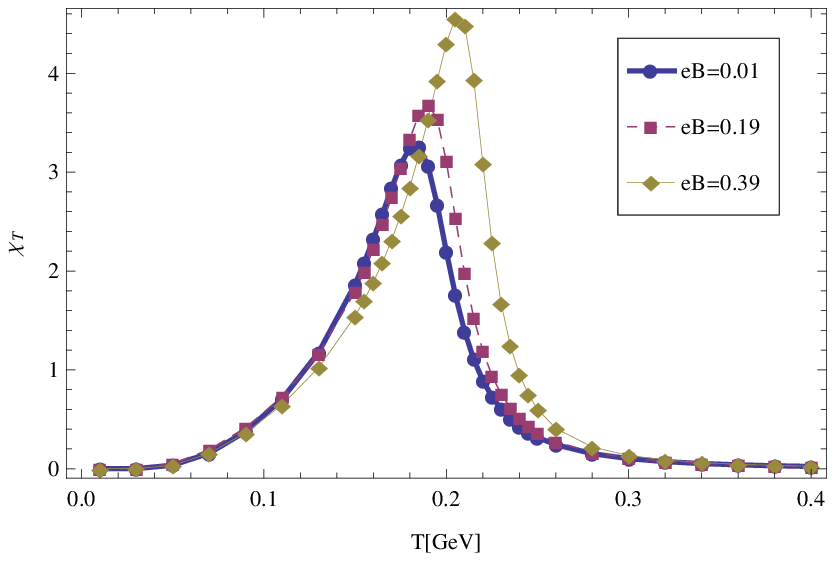}
\caption{The temperature dependance of $\chi_T$ with different $eB$.\label{ti}}
\end{figure}
\begin{figure}
\centering
\includegraphics[width=3in]{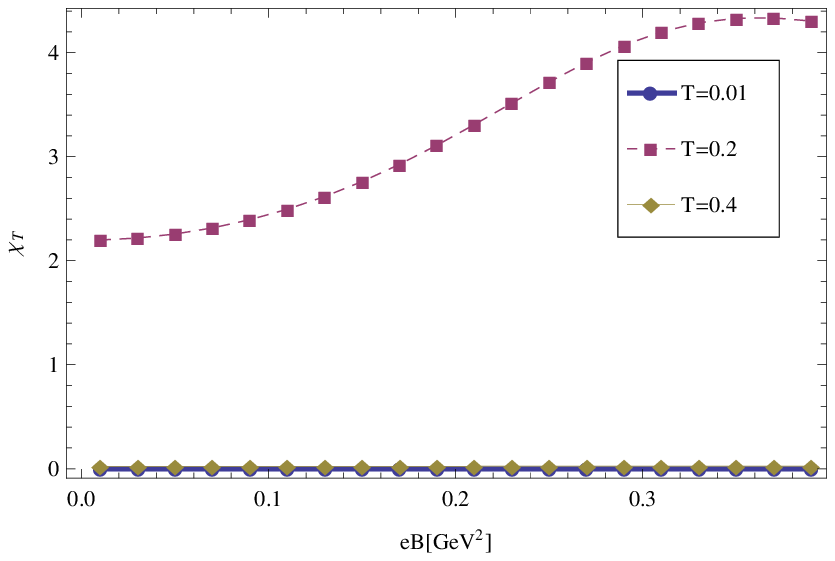}
\caption{The $eB$ dependance of $\chi_T$ with different temperature.\label{tii}}
\end{figure}
\begin{figure}
\centering
\includegraphics[width=3in]{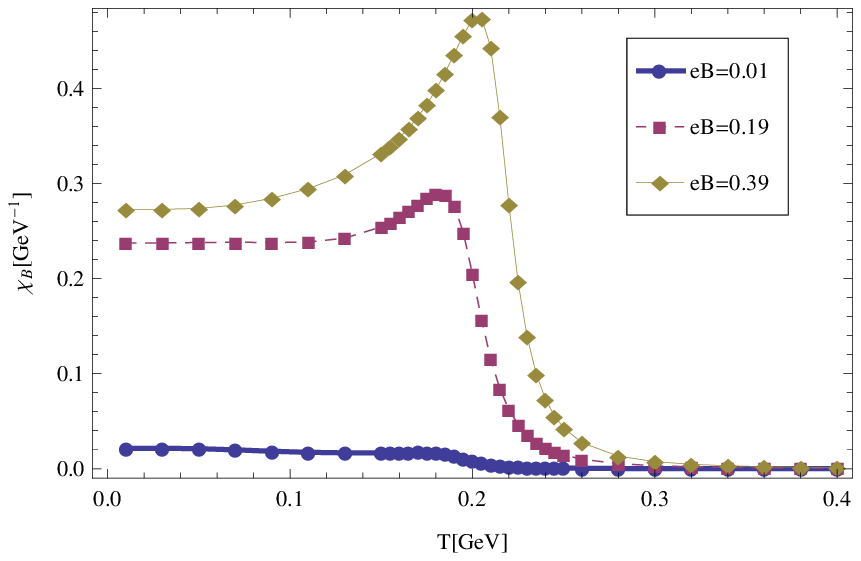}
\caption{The temperature dependance of $\chi_B$ with different $eB$.\label{bi}}
\end{figure}
\begin{figure}
\centering
\includegraphics[width=3in]{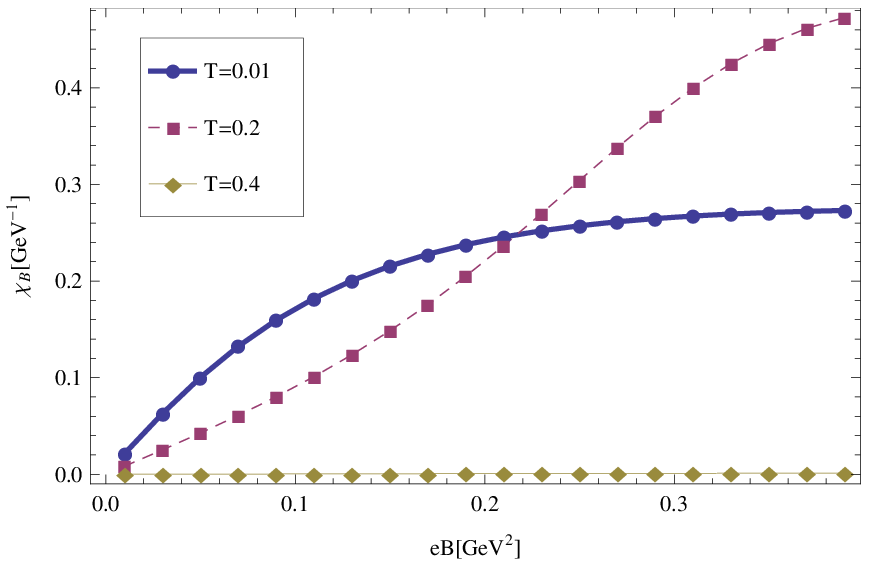}
\caption{The $eB$ dependance of $\chi_B$ with different temperature.\label{bii}}
\end{figure}
The current mass is fixed to $m=5.5\text{MeV}$, and the corresponding temperature or magnetic field dependance of susceptibilities are shown in Figs. \ref{ci}-\ref{bii}, in Figs. \ref{cii} and \ref{tii}, there seems to have anomalies, because the $T=0.2\text{GeV}$ curves in both figures have distinct values with $T=0.01\text{GeV}$ and $T=0.4\text{GeV}$ curves, this phenomenon is explainable, take Fig. \ref{cii} for example, making a cross-over analysis with Fig. \ref{ci}, in Fig. \ref{ci} the peak of $\chi_c$ is near $T=0.2\text{GeV}$, while at the both ends of $T$ ($T\to0$ and $T\to0.4$GeV), $\chi_c$ approaches small values, therefore in Fig. \ref{cii}, there comes the distinct difference between $T=0.2\text{GeV}$ and $T=0.01\text{GeV}$ $T=0.4\text{GeV}$ curves. In Figs. \ref{cii}, \ref{tii}, \ref{bii}, the peaks of $T=0.2$GeV curves tell us which value of $eB$ is the crossover points. And in these figures, we can use them to identify the crossover properties, for example, in Fig. \ref{cii}, the $T=0.2$GeV curve has no steep slopes around $eB=0.2$GeV$^2$, which means we can find out other crossover points at the nearby of $T=0.2$GeV with magnetic field not far from $eB=0.2$GeV$^2$.

From Figs. \ref{ci}, \ref{ti} and \ref{bi}, one can find the peaks are smooth, hence when the temperature increases from zero, the quark condensation will experience a crossover, and the crossover points is depending on external magnetic field, in these figures we can see when current mass has been considered, the crossover happens around $T=0.2\text{GeV}$, besides, in Fig. \ref{ci}, larger $eB$ will make the $T\text{-}\chi_c$ curve (take the peak as referent point) shift to right, in Fig. \ref{ti}, larger $eB$ not only makes $T\text{-}\chi_T$ curve shift to right but also makes the peak value increases, in Fig. \ref{bi}, larger $eB$ increases the values of $\chi_B$, but when $eB$ is small, the crossover behavior of $T\text{-}\chi_B$ is not so obvious, therefore when magnetic field is weak, $\chi_B$ is not a good parameter to study crossover or phase transition behaviors. In these three figures, we can see that with the same magnetic field, all these crossover points in different susceptibilities are basically consistent, the consistency comes from the property of $(1+\chi_c)$, because from Eqs. (\ref{chit}) and (\ref{chib}), they are all $(1+\chi_c)$-relevant.

\begin{figure}
\centering
\includegraphics[width=3in]{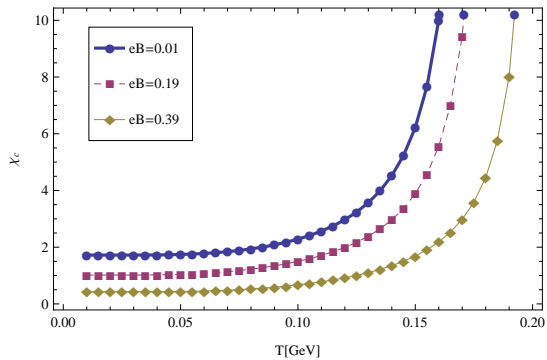}
\caption{The temperature dependance of $\chi_c$ with different $eB$ when $m=0$.\label{ciii}}
\end{figure}

\begin{figure}
\centering
\includegraphics[width=3in]{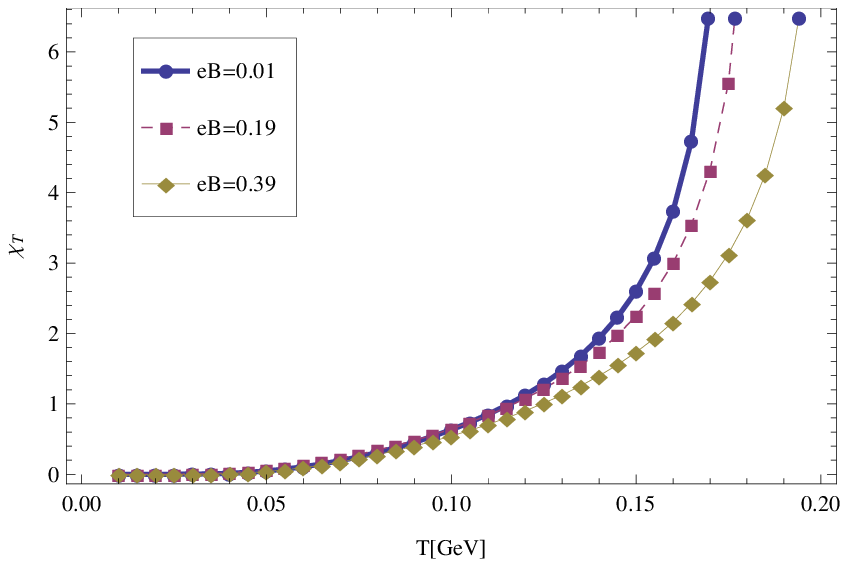}
\caption{The temperature dependance of $\chi_T$ with different $eB$ when $m=0$.\label{tiii}}
\end{figure}

\begin{figure}
\centering
\includegraphics[width=3in]{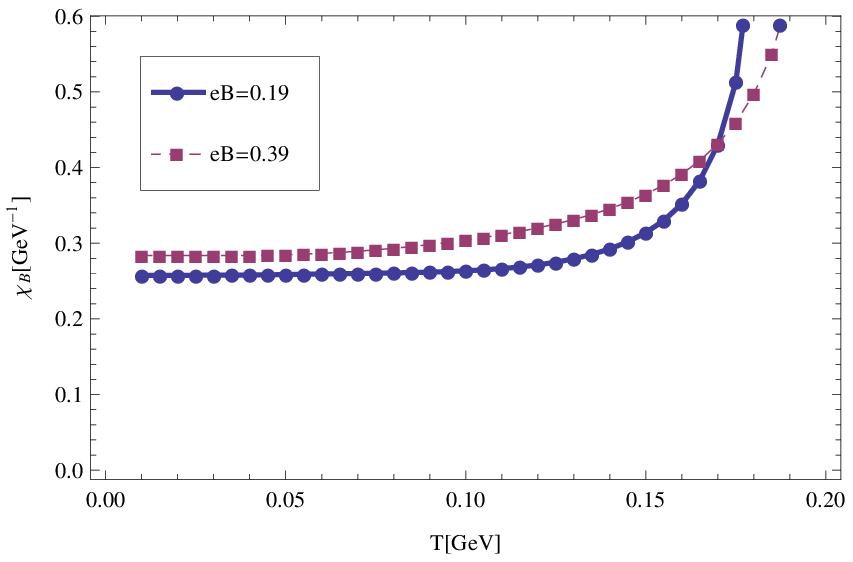}
\caption{The temperature dependance of $\chi_B$ with different $eB$ when $m=0$.\label{biii}}
\end{figure}

For comparison, we let current mass $m$ be zero, and draw the corresponding $\chi\text{-}T$ curves with different magnetic field, they are shown in Figs. \ref{ciii}, Fig. \ref{tiii} and Fig. \ref{biii}.
\begin{figure}
\centering
\includegraphics[width=3in]{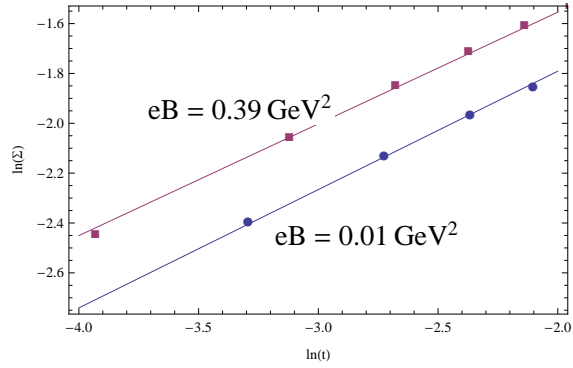}
\caption{The critical coefficient $\ln(\sigma)\text{-}\ln(t)$ relation when $m=0$, the slope is $\alpha$.\label{ccm}}
\end{figure}
\begin{figure}
\centering
\includegraphics[width=3in]{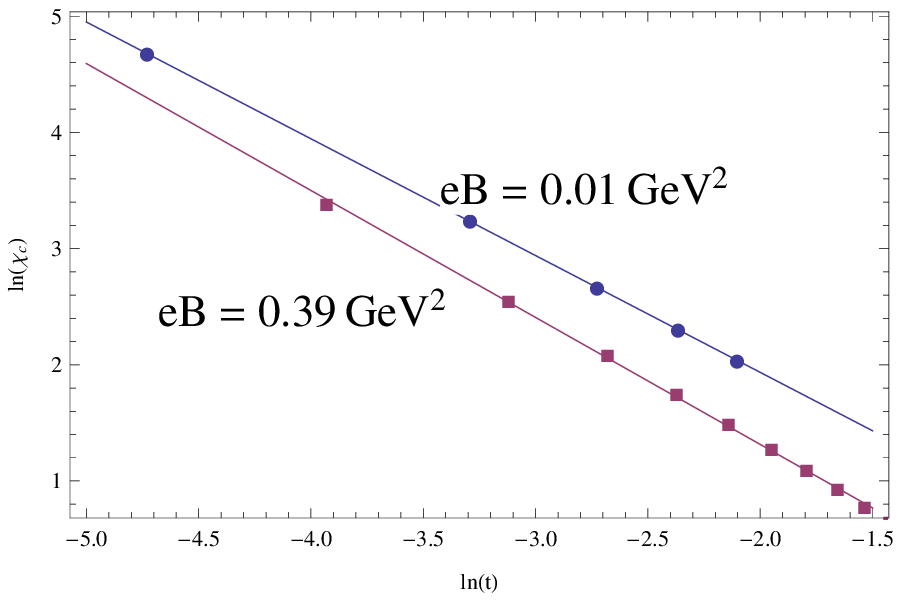}
\caption{The critical coefficient $\ln(\chi_c)\text{-}\ln(t)$ relation when $m=0$, the slope is $\gamma_c$.\label{ccc}}
\end{figure}
\begin{figure}
\centering
\includegraphics[width=3in]{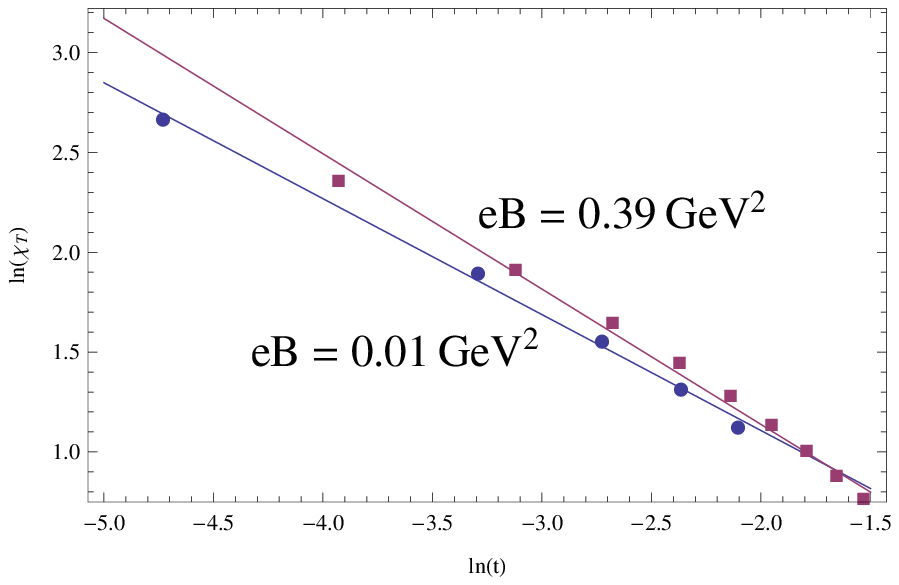}
\caption{The critical coefficient $\ln(\chi_T)\text{-}\ln(t)$ relation when $m=0$, the slope is $\gamma_T$.\label{cct}}
\end{figure}
\begin{figure}
\centering
\includegraphics[width=3in]{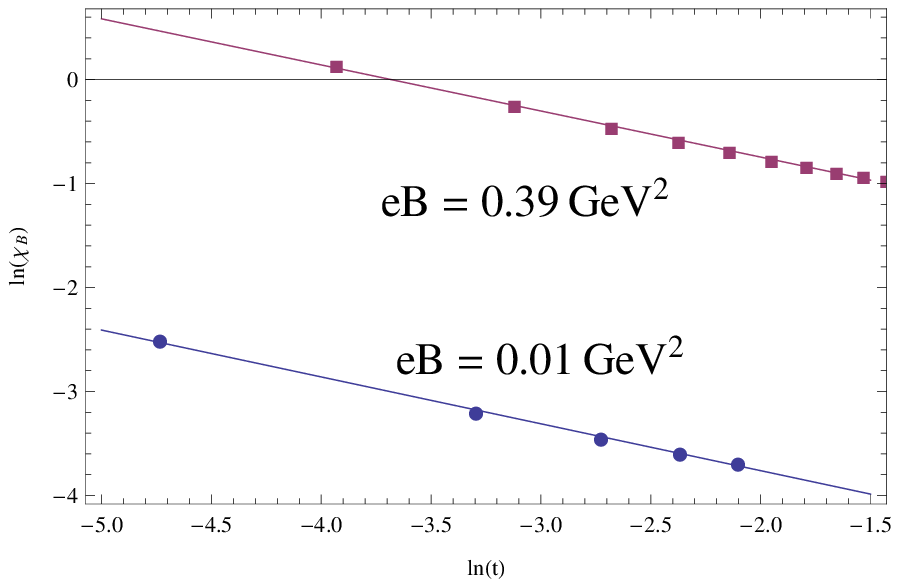}
\caption{The critical coefficient $\ln(\chi_B)\text{-}\ln(t)$ relation when $m=0$, the slope is $\gamma_B$.\label{ccb}}
\end{figure}

It is clear that when temperature approaches $0.2\text{GeV}$, all of the three susceptibilities approaches infinity, from Eq. (\ref{fm}) and Eq. (\ref{chic}), one can find out a reasonable explanation, because when $T$ crosses a critical point in the chiral limit, the dynamic mass $\sigma$ is zero, chiral symmetry is restored, this causes $f_m(0,0,T,eB)$ being zero, which leads to the infinity of $\chi_c$. For the other two susceptibilities $\chi_T$ and $\chi_B$, although $f_T(0,\sigma,T,eB)$ and $f_B(0,\sigma,T,eB)$ approach zero, but apparently the approaching rate is lower than $f_m(0,\sigma,T,eB)$, this ensures them also have infinity limits. After all, when $m=0$, there is no necessary to study the susceptibilities' properties at the $T>T_c$ area, because they are all infinities. These properties ensure us at chiral limit it is a second order phase transition. But one thing deserves mention here, at chiral limit, $T=0.2$GeV is not the upper limit of phase transition, although from Figs. \ref{ciii}, Fig. \ref{tiii} and Fig. \ref{biii} it seems so, but from Fig. \ref{ii}, we can see that bigger $eB$ will linearly increase $T_c$, eventually $T_c$ can exceed $0.2$GeV with no upper limit.

In order to study the second order phase transition, we also have studied the critical coefficients under chiral limit ($m=0$), near the phase transition point ($T$ is near and small than $T_c$), the definition of these coefficients are
\begin{equation}
t=1-\frac{T}{T_c},
\end{equation}
\begin{equation}
\sigma\sim t^\alpha,\quad\chi_c\sim t^{-\gamma_c},\quad\chi_T\sim t^{-\gamma_T},\quad\chi_B\sim t^{-\gamma_B}.
\end{equation}
$\alpha$, $\gamma_c$, $\gamma_T$ and $\gamma_B$ are the critical coefficients that we are going to find out, for calculating $\alpha$, on can draw the $\ln(\sigma)$-$\ln(t)$ diagram and make a linear fitting, the slope of linear fitting line is $\alpha$. The linear fittings are shown in Figs. \ref{ccm}, \ref{ccc}, \ref{cct}, \ref{ccb}, they are the logarithmic relations between these critical coefficients and $t$, in these figures we have taken two different magnetic fields for comparison. In Table \ref{tab}, we give the four kind of critical coefficients with different magnetic field.
\begin{table}
\caption{Critical coefficients with different $eB$}
\centering
\begin{tabular}{|c|c|c|c|c|}
\hline\hline
$eB$&$\alpha$&$\gamma_c$&$\gamma_T$&$\gamma_B$\\
\hline
0.01 & 0.47 & 1.00 & 0.58 & 0.45\\
\hline
0.05&0.47&1.00&0.59&0.44\\
\hline
0.11&0.45&1.01&0.63&0.42\\
\hline
0.21&0.47&1.02&0.60&0.45\\
\hline
0.31&0.47&1.04&0.60&0.46\\
\hline
\end{tabular}
\label{tab}
\end{table}

\section{Summary and Conclusions}
In this article, we have developed a new method to study the NJL gap equation with external magnetic field, this method is consistent with other method, and it is also compatible with more complicate equations such as Dyson-Schwinger equation, beside, this method is particularly convenient for dealing with constant magnetic field problems, while for a general external electromagnetic field the proper time method is still a better way. In our work, we have successfully repeated other researchers' works, these results are shown in Fig. \ref{i}-\ref{iv}, from these figures we know the 'Magnetic Catalysis' is not only enhance the QCD condensation but also increase the value of critical temperature. This is reasonable, because phenomenologically the external magnetic field restricts the movement modes of charge particles, gives more chance for any two particles to pair up, hence the QCD condensation increases. We have also calculated various susceptibilities, and all these susceptibilities imply that in chiral limit, the condensation has the second order chiral phase transition at finite temperature, while in non chiral limit, even the current mass is tiny, the condensation only experiences crossover. In this article we do not study chemical potential, because with chemical potential, there is a problem of chemical potential feed back, more than that, the presence of magnetic field will entangle the feed back, one can refer to the appendix for a simple impression, rigorous study of this problem is left in proceeding articles. In the chiral limit, the Wigner solution is trivial ($\sigma=0$), the susceptibilities of Wigner solution is also trivial $\chi_c=0$ (in Eq. (\ref{chic}), let $\sigma\to0$ first, ant then let $m\to0$, if one changes the limitation sequence, the result is infinity), hence we do not discuss in this paper.

\appendix
\section{A New Method Different from Schwinger Proper Time}
For simplicity, $\hat\Pi^f_\mu$ is wrote as $\hat\Pi_\mu$ in this appendix. Unlike free fermion propagator, $\hat\Pi_1$ and $\hat\Pi_2$ are not commutable, hence it is not possible to find a representation in which all four $\hat\Pi_\mu$'s eigenstates exist simultaneously, at least the eigenstates of $\hat\Pi_1$ and $\hat\Pi_2$ can not exist simultaneously. But for the denominator of fermion propagator, we do not need to find the eigenstates for all $\hat\Pi_\mu$, instead we turn to find the eigenstates of $\hat\Pi^2$.

Temporarily, we define the eigenstate of $\hat\Pi^2$ as $|\Pi^2_\perp,\Pi_0,\Pi_3\rangle$, in which $\Pi^2_\perp$ represents the eigenvalue of $(\hat\Pi_1^2+\hat\Pi_2^2)$, therefore $\hat\Pi^2|\Pi^2_\perp,\Pi_0,\Pi_3\rangle=(\Pi_0^2-\Pi^2_\perp-\Pi_3^2)|\Pi^2_\perp,\Pi_0,\Pi_3\rangle$. From the definition, we can see that the eigenstate only have three degrees of freedom, but to completely describe a particle there needs four degrees of freedom, therefore, in the following article, we are going to find out the rest degree of freedom. Besides, the operators $\hat\Pi_0$ and $\hat\Pi_3$ are actually $\hat p_0$ and $\hat p_3$, so the eigenvalues $\Pi_0$ and $\Pi_3$ are equivalent to $p_0$ and $p_3$.

For the eigenvalue of $\Pi_0$ and $\Pi_3$, their values vary continuously from $-\infty$ to $+\infty$, while the property of $\Pi^2_\perp$ are not identified yet. Because the operators $\hat\Pi^2_\perp$, $\hat\Pi_0$ and $\hat\Pi_3$ are commutable with each other, we can rewrite the eigenstate in a form of tensor product, $|\Pi^2_\perp,\Pi_0,\Pi_3\rangle=|\Pi_0\rangle_0\otimes|\Pi^2_\perp\rangle_{12}\otimes|\Pi_3\rangle_3$, which allows us to study the property of $\Pi^2_\perp$ separately. In the coordinates representation, $|\Pi^2_\perp\rangle_{12}$ lies in the $|x^1\rangle_1\otimes|x^2\rangle_2$ space, a general idea is to find the expression of $_{12}\langle x^1,x^2|\Pi^2_\perp\rangle_{12}$ through differential equation, but here we are not going to directly deduce the expression for $_{12}\langle x^1,x^2|\Pi^2_\perp\rangle_{12}$, instead we try to find an intermediate expression by introducing a new representation beyond the coordinates representation and momentum representation.

In the two dimension Hilbert space which in the language of coordinate representation is the tensor space on the basis of $|x^1\rangle_1\otimes|x^2\rangle_2$, we can reorganize the basis tensor through eigenstates of $\hat\Pi_1$. Because $\hat\Pi_1=\hat p_1-e\hat A_1=-\hat p^1+\frac{q_feB}{2}\hat x^2$, and the momentum operator $\hat p^1$ and coordinate operator $\hat x^2$ operate on different Hilbert spaces ($|x^1\rangle_1$ and $|x^2\rangle_2$ separately), then we can define $(\hat\Pi_1),\hat p^1$ as the complete operators that cover the whole $|x^1\rangle_1\otimes|x^2\rangle_2$ space, their eigenstates is
\begin{equation}
|\Pi_1,p\rangle=\sqrt{\frac{2}{|q_f|eB}}|p\rangle_1\otimes|x_{\Pi_1}\rangle_2,\qquad x_{\Pi_1}=\frac{2}{q_feB}(p+\Pi_1).\label{eigenstate}
\end{equation}
Here the eigenstate $|p\rangle_1$ is for $\hat p^1$ rather than $\hat p_1$, $\hat p^1|p\rangle_1=p|p\rangle_1$, $\hat p_1|p\rangle_1=-p|p\rangle_1$, one must caution the sign problem, actually in the context of this article, all the eigentstates of coordinates and momentum are eignestates of $\hat x^\mu$ or $\hat p^\mu$. In Eq. (\ref{eigenstate}), the factor $\sqrt{\frac{2}{|q_f|eB}}$ is used to normalize the basis tensors, hence we have the normalized, complete and orthogonal relations,
\begin{equation}
\langle\Pi'_1,p'|\Pi_1,p\rangle=\delta(\Pi-\Pi')\delta(p-p'),
\end{equation}
\begin{equation}
\int d\Pi_1dp\,|\Pi_1,p\rangle\langle\Pi_1,p|=1\otimes1,
\end{equation}
and by employing the equations $\hat p^2|x^2\rangle_2=i\frac{\partial}{\partial x^2}|x^2\rangle_2$ and $\hat x^1|p^1\rangle_1=-i\frac{\partial}{\partial p^1}|p^1\rangle_1$, we have
\begin{equation}
\hat\Pi_2|\Pi_1,p\rangle=-iq_feB\bigg(\frac{\partial}{\partial\Pi_1}-\frac{1}{2}\frac{\partial}{\partial p}\bigg)|\Pi_1,p\rangle.\label{partial}
\end{equation}

Now with this preparation of eigenstates $|\Pi_1,p\rangle$, we are able to establish a differential equation for $|\Pi^2_\perp\rangle_{12}$. For operator $\hat\Pi^2_\perp$, there is
\begin{equation}
\langle\Pi_1,p|\hat\Pi^2_\perp|\Pi^2_\perp\rangle=\Pi^2_\perp\langle\Pi_1,p|\Pi^2_\perp\rangle,\label{eq1}
\end{equation}
while $\hat\Pi^2_\perp=\hat\Pi^2_1+\hat\Pi^2_2$, with Eq. (\ref{partial}), there is also
\begin{equation}
\langle\Pi_1,p|\hat\Pi^2_\perp|\Pi^2_\perp\rangle=\Pi^2_1\langle\Pi_1,p|\Pi^2_\perp\rangle-q_f^2e^2B^2\bigg(\frac{\partial}{\partial\Pi_1}-\frac{1}{2}\frac{\partial}{\partial p}\bigg)^2\langle\Pi_1,p|\Pi^2_\perp\rangle,\label{eq2}
\end{equation}
put Eq. (\ref{eq1}) and Eq. (\ref{eq2}) together, we have the differential equation of $\langle\Pi_1,p|\Pi^2_\perp\rangle$, the variables are $\Pi_1$ and $p$,
\begin{equation}
\Pi^2_\perp\langle\Pi_1,p|\Pi^2_\perp\rangle=\bigg(\Pi^2_1-q_f^2e^2B^2\frac{\partial^2}{\partial\Pi_1^2}+q_f^2e^2B^2\frac{\partial^2}{\partial\Pi_1\partial p}-\frac{q_f^2e^2B^2}{4}\frac{\partial^2}{\partial p^2}\bigg)\langle\Pi_1,p|\Pi^2_\perp\rangle.\label{eq3}
\end{equation}
If we assume $\langle\Pi_1,p|\Pi^2_\perp\rangle$ has the following form,
\begin{equation}
\langle\Pi_1,p|\Pi^2_\perp\rangle=ce^{a(p+\frac{\Pi_1}{2})}h(z),\qquad z=\sqrt{\frac{2}{|q_f|eB}}\Pi_1.\label{solution}
\end{equation}
$a$ is an arbitrary complex constant, then Eq. (\ref{eq3}) can be simplified to
\begin{equation}
\frac{d^2h(z)}{dz^2}+\bigg(n+\frac{1}{2}-\frac{z^2}{4}\bigg)h(z)=0,\qquad n=\frac{\Pi^2_\perp}{2|q_f|eB}-\frac{1}{2}.\label{eq4}
\end{equation}
Eq. (\ref{eq4}) corresponds to Weber differential equation, the solutions of this differential equation depend on the property of $n$, here for the physical purpose, we need the solutions be convergent when $z\to\pm\infty$, therefore this requirement constrains $n$ be a nonnegative integer, which also determines the property of $\Pi^2_\perp$ as $\Pi^2_\perp=(2n+1)|q_f|eB$, $n\in\{0,1,2,\cdots\}$. In the following article, we rewrite function $h(z)$ as $h_n(z)$ to show the $n$-dependance of the solution. The general expression of $h_n(z)$ is
\begin{equation}
h_n(z)=e^{-\frac{z^2}{4}}z^n\sum_{k=0}^{[n/2]}\frac{(-\frac{n}{2})_k(\frac{1-n}{2})_k}{k!}\bigg(-\frac{z^2}{2}\bigg)^{-k},
\end{equation}
where $[n/2]$ is a nearest integer function, and $(-\frac{n}{2})_k$, $(\frac{1-n}{2})_k$ obey the following rule, for an arbitrary real number $\lambda$, a positive integer $k$, there is
\begin{equation}
(\lambda)_0=1,\quad(\lambda)_k=\frac{\Gamma(\lambda+k)}{\Gamma(\lambda)}=(\lambda+k-1)(\lambda+k-2)\cdots(\lambda+1)\lambda.
\end{equation}

Now we have identified $\Pi^2_\perp$ and $h_n(\sqrt{\frac{2}{|q_f|eB}}\Pi_1)$, let's go back to Eq. (\ref{solution}) to identify parameters $a$ and $c$. For the same physical consideration, we need $\langle\Pi_1,p|\Pi^2_\perp\rangle$ be convergent when $p,\Pi_1\to\pm\infty$, this leaves $a$ no other choice but a pure imaginary number, so we redefine $a$ as $ia$, now $a$ is an arbitrary real number, it represents the a hidden degree of freedom, hence in the following discussion we correct the eigenstate $|\Pi^2_\perp\rangle_{12}$ to $|n,a\rangle$.

$c$ is a normalization factor, to identify $c$'s value, we shall give the orthogonal and complete relations of $h_n(z)$ first,
\begin{equation}
\int h_m(z)h_n(z)\,dz=n!\sqrt{2\pi}\delta_{mn},
\end{equation}
\begin{equation}
\delta(x-y)=\sum_{n=0}^{+\infty}\frac{1}{n!\sqrt{2\pi}}h_n(x)h_n(y).
\end{equation}
In order to normalize $|n,a\rangle$, it is reasonable to assume $c$ is $n$-dependant, let $c\to c_n$,
\begin{equation}
\langle\Pi_1,p|n,a\rangle=c_ne^{ia(p+\frac{\Pi_1}{2})}h_n(\sqrt{\frac{2}{|q_f|eB}}\Pi_1),\quad c_n=\bigg(\frac{1}{n!2\pi\sqrt{|q_f|eB\pi}}\bigg)^\frac{1}{2}.
\end{equation}
Therefore eigenstate $|n,a\rangle$ has the following normalized and orthogonal relation,
\begin{equation}
\langle m,a'|n,a\rangle=\int d\Pi_1dp\,\langle m,a'|\Pi_1,p\rangle\langle\Pi_1,p|n,a\rangle=\delta_{mn}\delta(a-a').
\end{equation}

A short summary, for operators $(\hat\Pi^2_\perp,\hat\Pi_0,\hat\Pi_3)$, we can use their eigenstates $|\Pi_0,\Pi^3;n,a\rangle=|\Pi_0\rangle_0\otimes|n,a\rangle_{12}\otimes|\Pi_3\rangle_3$ as a set of complete basis tensors in four dimension Hilbert space.

Normally, we need to calculate $\Tr\hat S$, for example in Eq. (\ref{gap}), the operator $\Tr$ is representation irrelevant, hence with constant external magnetic field, $\Tr\hat S$ can be treated as
\begin{eqnarray}
\Tr\hat S&=&\int d\Pi_0d\Pi_3\int da\sum_{n=0}^{+\infty}\langle\Pi_0,\Pi_3;n,a|\tr\hat S|\Pi_0,\Pi_3;n,a\rangle\nonumber\\&=&2\sigma\int d\Pi_0d\Pi_3\,\langle\Pi_0,\Pi_3|\Pi_0,\Pi_3\rangle\sum_{n=0}^{+\infty}\frac{(2-\delta_{0n})}{\Pi_0^2-2n|q_f|eB-\Pi_3^2-\sigma^2}\int da\,\langle n,a|n,a\rangle.\label{trace}
\end{eqnarray}

In order to cancel $\int d^4x$ in LHS of Eq. (\ref{gap}), $\int d^4x$ must be extracted from Eq. (\ref{trace}), we can employ the following relations to serve this purpose,
\begin{equation}
\langle\Pi_0,\Pi_3|\Pi_0,\Pi_3\rangle=\frac{1}{(2\pi)^2}\int dx_0dx_3,
\end{equation}
\begin{equation}
\int da\,\langle n,a|n,a\rangle=\frac{|q_f|eB}{2\pi}\int dx_1dx_2,
\end{equation}
replacing them into Eq. (\ref{trace}), and let $\Pi_0$ have a Wick rotation, it will be simplified to
\begin{eqnarray}
\Tr\hat S&=&-i\frac{|q_f|eB\sigma}{\pi}\int d^4x\int\frac{d\Pi_0d\Pi_3}{(2\Pi)^2}\sum_{n=0}^{+\infty}\frac{(2-\delta_{0n})}{\Pi_0^2+2n|q_f|eB+\Pi_3^2+\sigma^2}\nonumber\\&=&-i\frac{|q_f|eB\sigma}{\pi}\int d^4x\int\frac{d\Pi_0d\Pi_3}{(2\Pi)^2}\sum_{n=0}^{+\infty}(2-\delta_{0n})\int_0^{+\infty}e^{-(\Pi_0^2+2n|q_f|eB+\Pi_3^2+\sigma^2)s}\,ds\nonumber\\&=&-i\frac{|q_f|eB\sigma}{4\pi^2}
\int d^4x\int_0^{+\infty}\frac{e^{-\sigma^2s}}{s}\coth(|q_f|eBs)\,ds.\label{trace2}
\end{eqnarray}

In this appendix, we have developed a new method that is different from Schwinger proper time method. The original idea of developing this new method is to evaluate how much influence does the infinitesimal imaginary term of fermion propagator has to the final results. Because in quantum field theory, the infinitesimal imaginary term (which always presents in the form of $i\varepsilon$) does not always play a role as pointer of integral path, in some case it can cause remarkable adjustment to the calculations, for example, in the case of zero temperature and finite chemical potential $\mu$, in Minkowski space \cite{alan,shu}, the fermion propagator is
\begin{equation}
S(k,\mu)=\frac{/\kern-0.4em\tilde k+m}{{\tilde k}^2-m^2+i\varepsilon(k_0+\mu)\sgn k_0},
\end{equation}
whose infinitesimal imaginary term is $\mu$-dependent, then after the Wick rotation of $p_0$, there will be nonzero residue counted in. Out of the consideration that the influence of a external magnetic field might affect infinitesimal imaginary term, therefor a thorough method is needed. For a free fermion propagator without any external field, the infinitesimal imaginary term is $i\varepsilon$, this is well-known in quantum field theory, but when external fields or external elements interfere, it is not safe to claim that the imaginary term is $i\varepsilon$, a convincible example is the case that the system has finite chemical potential, in Minkowski space, the infinitesimal imaginary term is total chemical potential dependance, therefore one need to properly deduce the the imaginary term with caution, generally there have two methods to deduce the fermion propagator with imaginary term, one is canonical quantization, the other is path integral. Here we introduced a convenient trick for path integral~\cite{mark,peskin}, the definition of partition function of quantum field theory is
\begin{equation}
Z=\left\langle 0\left|T\exp\bigg\{-i\int_{-\infty}^{+\infty}\hat H\,dt\bigg\}\right|0\right\rangle.
\end{equation}
Now introducing a factor $(1-i\eta)$ to change the expression of partition function, which will derive a $\eta$-dependent Lagrangian,
\begin{eqnarray}
Z&=&\lim_{\eta\to0^+}\left\langle 0\left|T\exp\bigg\{-i(1-i\eta)\int_{-\infty}^{+\infty}\hat H\,dt\bigg\}\right|0\right\rangle\nonumber\\
&=&\lim_{\eta\to0^+}\int d\psi\,\left\langle\psi\left|T\exp\bigg\{-i(1-i\eta)\int_{-\infty}^{+\infty}\hat H\,dt\bigg\}\right|\psi\right\rangle\nonumber\\
&=&\lim_{\eta\to0^+}\int D\bar\psi D\psi\,e^{i\int dx\,\mathcal{L}_\eta}.\label{partition}
\end{eqnarray}
In NJL model with external magnetic field, comparing with Eq. (\ref{propagator}), the improved fermion propagator is
\begin{equation}
\hat S=\frac{/\kern-0.6em\Pi+\sigma}{(/\kern-0.6em\Pi)^2-\sigma^2+iO(\eta)},\qquad O(\eta)=\eta(|\vec\Pi|^2+\sigma^2-eB\sigma^{12}).
\end{equation}
From previous discussion, we can see that $\Pi^2_\perp$ is quantized by magnetic field to $(2n+1)eB$, hence the imaginary term $O(\eta)$ is permanently positive, it is equivalent to $i\varepsilon$, the influence is trivial. But for the finite chemical potential case, one can prove that the tricks do give the right fermion propagator with $\mu$.

So far, we have used this new method to deduce the gap equations such in NJL model with magnetic field. In fact these gap equations are the same as the ones  deduced through Schwinger proper time method, the difference is the process, as mentioned before, this new method is developed to evaluate the infinitesimal imaginary term of fermion propagator, and of cause, the influence from imaginary term is trivial, but anyway, it deserves a mention how this been proved. This method is very convenient to study NJL problems with external magnetic field, because one does not need to find expression for $\langle x|\hat S|y\rangle$ at the first place, and it is also qualified for further study such as Schwinger-Dyson equations, but unfortunately this method can not simplify the complexity neither when higher order contributions such as fermion loop are considered.

\section{Deduction and Discussion of two flavor NJL gap equation with magnetic field}
With Eq. (\ref{L}), we are able to write the free energy as
\begin{equation}
\mathcal{F}=\frac{N_c}{2G}(\sigma^2+\pi^2)+N_ci\Tr_f\ln[i/\kern-0.5em\partial+e/\kern-0.65em A\otimes Q-(\sigma+i\gamma^5\otimes\vec\pi\cdot\vec \tau)],\label{free}
\end{equation}
the lower index '$f$' of $\Tr_f$ means beside the trace of spinor space and integral of $4$ dimension coordinate or momentum space there also has a trace of flavor space.

From Eq. (\ref{free}) one can deduce the gap equation for $\vec\pi$,
\begin{equation}
\frac{\vec\pi}{G}\int d^4x=i\Tr_f[(i\gamma^5\otimes\vec \tau)\hat S].\label{cp}
\end{equation}
When $A_\mu\equiv0$, one can easily prove that $\vec\pi$ can have trivial solution $\vec\pi=0$, normally in NJL model, this solution is widely accepted, and most of the researches are study the gap equation of $\sigma$. But in our case here, there are two reasons that we can not directly employ this conclusion, firstly, from Eq. (\ref{propagator}) one can find in the denominator of $\hat S_f$, the non-commutable relation of $\Pi^f_1$ and $\Pi^f_2$ produces the $q_feB\sigma^{12}$ term in the spinor space, secondly, the difference of electric charges of $u$ $d$ quarks produces a non-identity matrix $Q$ in the flavor space, therefore it is hard to deduce the final gap equations for $\sigma$ and $\vec\pi$, and $\vec\pi$ might not have a trivial solution. In this appendix, we are going to prove that $\vec\pi$ still can have a trivial solution.

The complete propagator $\hat S$ in Eq. (\ref{cp}) is
\begin{equation}
\hat S=\frac{1}{/\kern-0.55em p+e/\kern-0.65em A\otimes Q-(\sigma+i\gamma^5\otimes\vec\pi\cdot\vec \tau)}.
\end{equation}
In this propagator, because of the existence of $e/\kern-0.65em A\otimes Q$, it is not easy to construct simple quadratic terms in the denominator.

Since our purpose is to prove $\vec\pi$ can have trivial solution, we assume it is, then replace $\vec\pi=0$ into Eq. (\ref{cp}), if the LHS and RHS of Eq. (\ref{cp}) are equal, our assumption is proved.

When $\vec\pi=0$, the LHS of Eq. (\ref{cp}) is $0$, now we are going to prove that the RHS of Eq. (\ref{cp}) is also $0$. For the propagator, there is
\begin{equation}
\hat S|_{\vec\pi=0}=\frac{1}{/\kern-0.55em p+e/\kern-0.65em A\otimes Q-\sigma}=\left(\begin{array}{cc}\frac{1}{/\kern-0.5em\hat\Pi^u-\sigma}&O\\O&\frac{1}{/\kern-0.5em\hat\Pi^d-\sigma}\end{array}\right)
=\left(\begin{array}{cc}\hat S_u&O\\O&\hat S_d\end{array}\right).
\end{equation}
Therefor, we have
\begin{equation}
(i\gamma^5\otimes \tau^3)(\hat S|_{\vec\pi=0})=\left(\begin{array}{cc}i\gamma^5\hat S_u&O\\O&-i\gamma^5\hat S_d\end{array}\right),
\end{equation}
replacing it into the RHS of Eq. (\ref{cp}),
\begin{equation}
\Tr_f[(i\gamma^5\otimes \tau^3)(\hat S|_{\vec\pi=0})]=\Tr(i\gamma^5\hat S_u)-\Tr(i\gamma^5\hat S_d)=0.
\end{equation}
As for $\tau^{1,2}$ of $\vec \tau$, there definitely has
\begin{equation}
\Tr_f[(i\gamma^5\otimes \tau^{1,2})(\hat S|_{\vec\pi=0})]=0.
\end{equation}
Now we have proved that with the presence of external magnetic field, we still can find trivial solution for $\vec\pi$. In fact this is rational, because $i\gamma^5\otimes\vec\pi\vec\tau$ in the denominator of complete fermion propagator violates parity, hence in many papers $\vec\pi$ is set as $0$~\cite{klevansky,buballa}.

Through this conclusion, now we are able to simplify the gap equation of $\sigma$,
\begin{equation}
\frac{\sigma}{G}\int d^4x=i\Tr_f(\hat S)=i\Tr_f(\hat S|_{\vec\pi=0})=i\sum_f\Tr(\hat S_f).
\end{equation}

Generally speaking, $\sigma$ should be a $(4\times4)\otimes(2\times2)$ matrix, but from previous discussion we know that the $(2\times2)$ part (flavor space) is eventually simplified by the $\Tr_f$ operator and $\vec\pi=0$, if one demands $\vec\pi\neq0$, then the gap equations of $\sigma$ and $\vec\pi$ will definitely not be such simple. On the other hand the $(4\times4)$ part (spinor space) is simplified to scaler at the beginning in Eq. (\ref{L}), because $\sigma$ is introduced here as an auxiliary scaler field.

Actually in NJL model, it is not obvious that the dynamic mass $\Sigma$ should be a $(4\times4)$ matrix (here for convenience we only consider one flavor NJL model, the flavor space is neglected), because either mean field approximation or auxiliary field method, it provides only a scaler rather than a matrix to the dynamic mass (or non-approximate fermion self energy). While in Ref. \cite{asakawa}'s work, the authors had proposed a method to find out the matrix form of self energy, they used Fierz transformation to identify the matrix structure. But in a matter of fact, the result is trivial as long as chemical potential is not involved.

Instead of employing the method used in Ref. \cite{asakawa}, here we use Dyson-Schwiger equation with contact interaction model to clarify this problem, the Dyson-Schwinger equation of fermion self energy is
\begin{equation}
-i\Sigma(x,y)=g^2\int dz_1dz_2\,\gamma^\mu D_{\mu\nu}(x,z_1)S(x,z_2)\Gamma^\nu(z_1;z_2,y).\label{ds}
\end{equation}
For the contact interaction model, the gluon propagator and dressed vertex are simplified into
\begin{equation}
D_{\mu\nu}(x,z_1)=\lambda g_{\mu\nu}\delta(x-z_1),\quad\Gamma^\nu(z_1;z_2,y)=\gamma^\nu\delta(z_1-y)\delta(z_2-z_1),
\end{equation}
replacing them into Eq. (\ref{ds}),
\begin{equation}
\Sigma=i\lambda g^2\langle x|\gamma^\mu\frac{1}{\hat S_0^{-1}-\Sigma}\gamma_\mu|x\rangle.\label{dsc}
\end{equation}
In Eq. (\ref{dsc}), it is $\gamma^\mu\hat S\gamma_\mu$ rather than $\Tr\hat S$ in the NJL gap equation, this difference naturally leads to the assumption that $\Sigma$ is a $4\times4$ spinor space matrix as long as it complies with some specific symmetry. Normally $\Sigma$ should be $\Sigma=\sigma+a_\mu\gamma^\mu$, there is no $\gamma^5$ and $\gamma^5\gamma^\mu$ terms because they violate parity, there is also no $\sigma^{\mu\nu}$ terms because if there was, in the LHS of Eq. (\ref{dsc}), the $\sigma^{\mu\nu}$ terms of $\Sigma$ can not find corresponding terms in the RHS because $\gamma^\mu\sigma^{\tau\nu}\gamma_\mu=0$. On the other hand, $a_\mu$ in $\Sigma$ can be absorbed by $p_\mu$, hence their contributions are trivial, this leaves only $\sigma$ contributes to the self energy $\Sigma$, therefore a $\Sigma$ can be simplified to a scaler.

In this article, although magnetic field has been involved, we can also prove that $a_\mu$ in $\Sigma=a_\mu\gamma^\mu+\sigma$ is trivial, and a scaler $\Sigma$ is the simplest and self-consistence solution for our magnetic field case. Assuming $\Sigma=a_\mu\gamma^\mu+\sigma$ (there still not has $\gamma^5$, $\gamma^5\gamma^\mu$, $\sigma^{\mu\nu}$ terms for the same reason as mentioned above), from Appendix A, we know
\begin{equation}
\int\langle x|\frac{1}{/\kern-0.6em\Pi-a_\mu\gamma^\mu-\sigma}|x\rangle\,d^4x=aI_4+b\sigma^{12}.
\end{equation}
And for the spinor matrices, there is
\begin{equation}
\gamma^\mu\sigma^{\tau\nu}\gamma_\mu=0,
\end{equation}
hence there is
\begin{equation}
\sigma\int d^4x=i\lambda g^2\gamma^\mu\bigg(\int\langle x|\hat S|x\rangle\,d^4x\bigg)\gamma_\mu=4i\lambda g^2a,
\end{equation}
the LHS and the RHS are both scalers, the assumption of scaler $\Sigma$ is reasonable.

\section*{Acknowledgements}
This work is supported in part by the National Natural Science Foundation of China (under Grants No. 11275097 and No. 11475085, 11265017), the China Postdoctoral Science Foundation (under Grant No. 2014M561621), the Jiangsu Planned Projects for Postdoctoral Research Funds (under Grants No. 1401116C and No. 1402006C), the National Basic Research Program of China (under Grant No. 2012CB921504), and the Guizhou province outstanding youth science and technology talent cultivation object special funds (QKHRZ(2013)28).

\end{document}